# Quantum-electrodynamical approach to the Casimir force problem

Frédéric Schuller[1] and Renaud Savalle[2]

[1] Laboratoire de Physique des Lasers, UMR 7538 du CNRS, Université Paris 13, F-93430 Villetaneuse, France

[2] Observatoire de Paris, 5 Pl Jules Janssen, 92195 Meudon, France. Tel : +33 (0)1 45 07 75 86. Fax : +33 (0)9 58 41 31 41. E-mail : renaud.savalle@obspm.fr

Abstract

We derive the Casimir force expression from Maxwell's stress tensor by means of original quantum-electro-dynamical cavity modes. In contrast with similar calculations, our method is straightforward and does not rely on intricate mathematical extrapolation relations.

Introduction

The effect of the quantum nature of the electro-magnetic field manifests itself if retarded potentials between atoms and molecules are considered, as shown by Casimir and Polder [1]. An even more striking quantum effect with no classical analogue is the attractive force between perfectly reflecting parallel plates at zero temperature. For this effect Casimir derived in 1948 the expression [2]

$$(1) \quad F_z = \frac{\pi^2}{240}\frac{\hbar c}{a^4}$$

with the plates located at z = 0 and z = a.

In his derivation Casimir considers the zero point energy inside the cavity, and he obtains the force as represented by its derivative with respect to the distance a.

Here a difficulty appears given the fact that if the zero-point energy density

$$(2) \quad \varepsilon = \frac{1}{2}\hbar\omega_{\vec{k}}$$

is summed over all possible modes $\vec{k}$, an infinite meaningless result is obtained. Casimir has solved the problem by demonstrating how unphysical infinities disappear by introducing compensations from the surroundings outside the cavity. However, this delicate problem has given rise to numerous subsequent studies [3], most of them based on intricate mathematical extrapolation methods.

Another way, chosen by some authors for computing the Casimir force, consists in writing down an expression derived directly from Maxwell's stress tensor instead of differentiating the energy relation. Unfortunately this method does not remove the divergencies brought about by summation over modes so that here again convergence generation procedures are required. This approach has been initiated by Brown and Maclay [4] for the general case of finite temperatures. They use temperature Green's functions in a way appropriate to deal with divergencies Numerous other methods have been published since for which we refer to the book of K.A.Milton [3].

Here our aim is to show that in the special case of zero temperature the right result is obtained without the detour through the extensive temperature Green's function formalism [5]. In doing so, we evaluate Maxwell's stress tensor by means of specific quantum-electro-dynamical cavity modes which to our knowledge have not been considered so far in the literature in this context. The advantage of using these modes lies in the fact that they allow to perform a rigorous mathematical treatment, with at the end a clear identification of the infinity that has to be discarded as unphysical.

The cavity modes

We consider a cavity made of two plane parallel perfectly conducting metallic plates located at distances $z = 0$ and $z = a$ respectively. We write the electric field vector inside the cavity in the form

(3) $\hat{\vec{E}} = \vec{E}\left(\hat{a}e^{-i\omega t} + \hat{a}^+ e^{i\omega t}\right)$

with, for a given frequency, $\vec{E}$ a function of the space coordinates $\vec{r} = (x, y, z)$ and $\hat{a}$, $\hat{a}^+$ the usual annihilation and creation operators respectively. Introducing the expectation value of the quantity $\hat{\vec{E}}^2$ in the vacuum state $|0\rangle$ we then have

(4) $\langle 0|\hat{\vec{E}}^2|0\rangle = \vec{E}^2 \langle 0|\hat{a}\hat{a}^+|0\rangle = \vec{E}^2$

Considering now the zero point energy density in the cavity as given by the expression

(5) $\frac{1}{2}\hbar\omega / V = \frac{1}{2}\frac{\hbar\omega}{L^2 a}$

where $L^2$ is the surface of the quare shaped plates, the quantity given by eq.(4) can be linked to the electro-magnetic energy density, i.e.twice its electric part, by means of the following relation:

(6) $2\varepsilon_0 \vec{E}^2 = \frac{1}{2}\frac{\hbar\omega_k}{L^2 a}$

Here we have added the index k as the corresponding wave number.

We now turn to the function $\vec{E}$ which for a given mode with wave vector $\vec{k}$ has to fullfill two conditions:

i) the zero charge condition inside the cavity expressed by the relation

(7) $\nabla \cdot \vec{E} = 0$

ii) the continuity condition of the parallel components at the boundaries, which for conducting plates amounts to setting

(8) $E_x = E_y = 0$ for $z = 0$, $z = a$.

A candidate obeying these conditions is given by the expression [8]

(9) $\begin{aligned} E_x &= A_x \cos(k_x x)\sin(k_y y)\sin(k_z z) \\ E_y &= A_y \sin(k_x x)\cos(k_y y)\sin(k_z z) \\ E_z &= A_z \sin(k_x x)\sin(k_y y)\cos(k_z z) \end{aligned} \qquad \vec{k} = \begin{pmatrix} k_x \\ k_y \\ k_z \end{pmatrix} = \begin{pmatrix} \dfrac{\pi n_x}{L} \\ \dfrac{\pi n_y}{L} \\ \dfrac{\pi n_z}{a} \end{pmatrix}$

with from eq.(7) the condition

(9a) $A_x k_x + A_y k_y + A_z k_z = 0$

Moreover, by averaging over space coordinates eq.(6) then takes the form

(10a) $\varepsilon_0 2 \times \dfrac{1}{8} A^2 = \dfrac{1}{2}\dfrac{\hbar \omega_k}{L^2 a}$

(10b) $A^2 = \dfrac{2}{\varepsilon_0}\dfrac{\hbar \omega_k}{L^2 a}$

with

(11) $A^2 = A_x^2 + A_y^2 + A_z^2$

The Casimir force

The force exerted by the cavity field on the conducting plates can be deduced from Maxwell's stress tensor

(12) $\sigma_{ij} = \varepsilon_0 E_i E_j + \dfrac{1}{\mu_0} B_i B_j - \dfrac{1}{2}\left(\varepsilon_0 \vec{E}^2 + \dfrac{1}{\mu_0}\vec{B}^2\right)\delta_{ij}$ with $i,j = x,y,z$

where the electric and magnetic field quantities refer to their values at the boundaries. For the electric field these are given by eq.(8) and (9). For the magnetic field they can be deduced from the relation

(13) $\vec{B} = -\dfrac{1}{i\omega} \nabla \times \vec{E}$

yielding the following results

(14) $B_z = 0$

(15) $\left(\nabla \times \vec{E}\right)_{x,y} = \begin{pmatrix} \left(A_z k_y - A_y k_z\right) \sin(k_x x) \cos(k_y y) \\ -\left(A_z k_x - A_x k_z\right) \cos(k_x x) \sin(k_y y) \end{pmatrix}$

Taking the square of the quantity given by eq.(13) and averaging over coordinates x,y, we then obtain after some algebra

(16) $\left\langle \vec{B}^2 \right\rangle_{xy} = \dfrac{1}{4} \dfrac{1}{\omega^2} \left(A_z^2 k^2 + A^2 k_z^2\right) = \dfrac{1}{4} \dfrac{1}{c^2} \left(A_z^2 + A^2 \dfrac{k_z^2}{k^2}\right)$

Where from eq.(9a) the relation

(17) $A_x k_x + A_y k_y = -A_z k_z$

has been used.

Replacing the quantities of eq.(12) by these averages together with, as a result of eq.(9),

(18) $\left\langle E_z^2 \right\rangle_{xy} = \left\langle \vec{E}^2 \right\rangle_{xy} = \dfrac{1}{4} A_z^2$

we arrive at the relation

(19) $\sigma_z = \varepsilon_0 E_z^2 - \dfrac{1}{2}\left(\varepsilon_0 \vec{E}^2 + \dfrac{1}{\mu_0} \vec{B}^2\right) = -\dfrac{1}{8} \varepsilon_0 A^2 \dfrac{k_z^2}{k^2}$

given the fact that the terms in $A_z^2$ cancel and recalling that $\dfrac{1}{c^2} = \mu_0 \varepsilon_0$.

With $A^2$ given by eq.(10b) we have

(20) $\sigma_z = \dfrac{\hbar c k}{L^2 a} \dfrac{k_z^2}{k^2}$

Setting

(21) $\kappa^2 = k_x^2 + k_y^2$

and using the explicit expressions of eq.(9) for $k_x, k_y$ together with

$k = \sqrt{\kappa^2 + k_z^2} = \sqrt{\kappa^2 + \left(\dfrac{n_z \pi}{a}\right)^2}$

we thus obtain

$$(22) \quad \sigma_z = -\frac{1}{4}\frac{\hbar c}{L^2 a}\frac{\left(\frac{n_z \pi}{a}\right)^2}{\sqrt{\kappa^2 + \left(\frac{n_z \pi}{a}\right)^2}}$$

We now have to sum over mode numbers $n_x, n_y, n_z$. Letting the auxiliary quantity L go to infinity we make the usual replacement

$$(23) \quad \sum_{n_x n_y} \to \frac{L^2}{\pi^2} dk_x dk_y = \frac{L^2}{\pi^2} d^2\kappa$$

This then yiedls for the force the expression:

$$(24) \quad F_z = -\frac{1}{a}\frac{\hbar c}{4\pi^2}\sum_n \int d^2\kappa \frac{\frac{n^2 \pi^2}{a^2}}{\sqrt{\kappa^2 + \left(\frac{n^2 \pi^2}{a^2}\right)}} \quad \text{(summing over all positive integers)}$$

This expression is highly divergent as it should, given the facts mentioned in the introduction. Therefore a convergence factor has to be inserted which, following Fierz [7], we choose to be $e^{-\lambda k}$.

Naturally the result will depend on the frequency cut-off involving the value of $\lambda$. The sum on the r.h.s. of eq.(24) can be done exactly as shown in the appendix, yielding the result

$$(25) \quad F_z = -\frac{\hbar c}{\pi^2}\lambda^{-4} - \frac{\hbar c}{a^4}\pi^2 \frac{6}{2\times 4!}B_4 \quad \text{with } B_4 = -\frac{1}{30} \text{ a Bernoulli number}$$

Hence the second term on the r.h.s. is equal to the Casimir force as given by eq.(1)

Discussion

In eq.(25) the infinity problem is represented by the first term. Since this term does not depend on the parameters of the system, i.e. of the distance a, it can safely be discarded as unphysical. This argument makes our method different from the numerous approaches which consider in detail compensation of singularities by various procedures.

Note that the expression (A1) represents twice (2 polarization states) the derivative with respect to the distance a of the corresponding scalar expression, provided however that the derivative of the convergence factor can be ignored. This latter important condition has been investigated in detail in ref. [8]

We finally mention that a general rule in quantum physics stipulates that the result of any actual measurement involving the zero-point energy should be proportional to $(\Delta l)^{-4}$ with $\Delta l$ a small characteristic length [9]. This is because the measurement reflects the average over a small but finite region in space. This is in accordance with the Casimir formula where $\Delta l$ can be identified with the distance a.

Conclusion

We present a derivation of the Casimir formula by evaluating Maxwell's stress tensor in terms of quantum electrodynamical cavity modes. The calculations involve only straightforward mathematical developments and convergence is achieved on purely physical grounds.

Bibliography

[1] H.G.B.Casimir and D.Polder, Phys.Rev.73 , 360, (1948)

[2] H.G.B.Casimir, Proc.Kon.Ned.Acad.Wetenschap **57**,61 (1948)

[3] K.A.Milton, The Casimir Effect. World Scientific,Singapore (2001)

[4] L.S.Brown and C.G.Maclay, Phys.Rev.184,1272, (1969)

[5] Landau and Lifshitz, Course of Theoretical Physics, Volume 9

[6] R. Loudon The Quantum Theory of Light, Clarendon Press Oxford (1973)

[7] M.Fierz, Helv.Phys.Acta, **33**,855 (1960)

[8] F.Schuller, Z. Naturforsch. 63a ,571, 2008

[9] J.J. Sakurai, Advanced Quantum Mechanics, Addison - Wesley (1967) p.33

Appendix

With k given by eq.(23) we obtain for $F_z$ the expression

$$(A1)\quad F_z = -\frac{1}{4\pi^2}\hbar c \frac{1}{a}\sum_n \int d^2\kappa \frac{\frac{n^2\pi^2}{a^2}}{\sqrt{\kappa^2+\frac{n^2\pi^2}{a^2}}} e^{-\lambda\sqrt{\kappa^2+\frac{n^2\pi^2}{a^2}}}$$

We evaluate this quantity by making the following substitutions [9] :

$$d^2\kappa = 2\pi\kappa d\kappa \quad ; \quad \kappa^2 = \frac{n^2\pi^2}{a^2}z$$

We thus obtain

(A2) $F_z = -\frac{\hbar c}{4\pi} \frac{1}{a} \sum_n \left(\frac{n^3 \pi^3}{a^3}\right) \int_0^\infty \frac{1}{\sqrt{z+1}} e^{-\lambda \frac{n\pi}{a}\sqrt{z+1}} dz$

The integral can be calculated exactly with the result

(A3) $F_z = -\frac{\hbar c}{2\pi} \frac{1}{a} \sum_n \frac{n^2 \pi^2}{a^2} \frac{1}{\lambda} e^{-\lambda \frac{n\pi}{a}}$

After generating the factor $n^2$ by deriving the exponential twice with respect to $\lambda$, the sum over n reduces to a geometric series. This yields the expression

(A4) $F_z = -\frac{\hbar c}{2\pi} \frac{1}{\lambda} \frac{\partial^2}{\partial \lambda^2} \frac{1}{1 - e^{-\lambda \frac{\pi}{a}}}$

Introducing Bernoulli numbers $B_h$ defined by

(A5) $\frac{-\lambda \frac{\pi}{a}}{e^{-\lambda \frac{\pi}{a}} - 1} = \frac{\lambda \frac{\pi}{a}}{1 - e^{-\lambda \frac{\pi}{a}}} = \sum_0^\infty \frac{B_h}{h!} (-1)^h \left(\frac{\pi}{a}\right)^h \lambda^h$

we expand the expression (A4) as follows:

(A6) $F_z = -\frac{\hbar c}{2\pi} \frac{1}{a} \sum_0^\infty \frac{B_h}{h!} (-1)^h \left(\frac{\pi}{a}\right)^{h-1} (h-1)(h-2) \lambda^{h-4}$

Letting $\lambda$ be an infinitesimal quantity, eq.(A6) reduces to

(A7) $F_z = -\frac{\hbar c}{\pi^2} \lambda^{-4} - \frac{\hbar c}{a^4} \pi^2 \frac{6}{2 \times 4!} B_4$

with $B_0 = 1$, $B_3 = 0$, $B_4 = -\frac{1}{30}$

We thus recover the result of eq.(25).